\documentclass[aps,reprint,article,floatfix,twocolumn,nofootinbib]{revtex4-2} 

\usepackage{longtable} 
\usepackage[dvipsnames]{xcolor}
\usepackage[utf8]{inputenc} 
\usepackage[T1]{fontenc} 
\usepackage{graphicx} 
\usepackage{amssymb,amsmath} 
\usepackage{amsfonts} 
\usepackage{units}
\usepackage{overpic} 
\usepackage{array} 
\usepackage{calc} 
\usepackage{relsize} 
\usepackage[english]{babel} 
\usepackage[savemem]{listings} 
\usepackage{wrapfig}
\usepackage{float} 
\usepackage{lmodern}
\usepackage{parskip}
\usepackage{subfigure} 
\usepackage{makeidx} 
\usepackage{enumerate}
\usepackage{soul}

\usepackage{multirow}
\usepackage{bbm} 
\usepackage{wasysym}

\usepackage[pdftex,
			pdfa,
			colorlinks,
			pdfpagelabels,
			pdfstartview = Fit,
			bookmarksopen = false,
			bookmarksnumbered = true,
			linkcolor = black,
			plainpages = false,
			hypertexnames = false,
			citecolor = black,
			urlcolor = blue]{hyperref} 

\newcommand{\1}{\begin{equation}}
\newcommand{\2}{\end{equation}}
\newcommand{\ea}{\begin{eqnarray}} 
\newcommand{\ee}{\end{eqnarray}}

\newcommand{\Sum}[2]{{\sum\limits_{#1}^{#2}}} 

\newcommand{\erw}[1]{\left\langle\, #1\,\right\rangle} 

\newcommand{\verti}[1]{\erw{\hat n_{{\bf k}\sigma}}} 

\newcommand{\bee}{\begin{eqnarray*}}
\newcommand{\eee}{\end{eqnarray*}}

\newcommand{\e}{{\rm e}}

\newcommand{\sa}{\left[ \begin{array} {c} }
\newcommand{\se}{\end{array}\right]}

\newcommand{\kb}{k_{\text{B}}}

\begin{document}
\title{An Introduction to Modeling Approaches of Active Matter}

\author{L.\ Hecht}
 \affiliation{ 
 	Institut für Physik kondensierter Materie, Technische Universität Darmstadt, Hochschulstr.\ 8, 64289 Darmstadt, Germany
 }
\author{J.\ C.\ Ureña}
 \affiliation{ 
 	Institut für Physik kondensierter Materie, Technische Universität Darmstadt, Hochschulstr.\ 8, 64289 Darmstadt, Germany
 }
\author{B.\ Liebchen}
 \email{liebchen@fkp.tu-darmstadt.de}
 \affiliation{ 
 	Institut für Physik kondensierter Materie, Technische Universität Darmstadt, Hochschulstr.\ 8, 64289 Darmstadt, Germany
 }

\date{\today}

\maketitle

\tableofcontents

\section{Introduction}

This article is based on lecture notes for the Marie Curie Training school ``Initial Training on Numerical Methods for Active Matter''. It provides an introductory overview of modeling approaches for active matter and is primarily targeted at PhD students (or other readers) who encounter some of these approaches for the first time. The aim of the article is to help put the described modeling approaches into perspective.

We begin with a brief discussion of the role of the solvent in (soft) active matter, which is followed by an introduction to ``dry particle-only models'', such as the active Brownian particle model, before coming to models for wet active matter, which (explicitly) include a solvent, and continuum descriptions for the collective behavior of many active particles.

\section{Wet and dry models: The role of the solvent in active matter} 

Models of active matter can be classified into ``dry'' and ``wet'' models. The former class of models involves only equations of motion for ``particles'' whereas the latter involves an explicit description of a solvent in addition to the embedded active particles, which ensures momentum conservation, as will be discussed below.

Dry models are naturally used to describe active systems which do not involve a liquid solvent, such as granular particles on vibrating plates \cite{Scholz_NatCom_2018,Scholz_NatCom_2018_2,Scholz_NewJPhys_2016,Lanoiselee_PhysRevE_2018,Kudrolli_PhysRevLett_2008,Walsh_SoftMatter_2017}, self-vibrating granular particles \cite{Dauchot_PhysRveLett_2019,Deblais_PhysRevLett_2018,Patterson_PhysRveLett_2017}, bacteria gliding on a rigid surface \cite{Wolgemuth_CurrBiol_2002}, flocks of birds, animal herds, swarms of locusts \cite{Toner_PhysRevE_1998,Buhl_Science_2006,Hemelrijk_InterFoc_2012,Ballerini_PNAS_2008,Bialek_PNAS_2012}, human crowds \cite{Silverberg_PhysRevLett_2013,Bain_Science_2019}, or flying drones \cite{Vasarhelyi_SciRob_2018}. However, they are also frequently used as simplified descriptions of active matter systems involving a solvent which is only effectively represented and commonly acts as a thermal bath leading to fluctuations in the equations of motion of the individual particles. In contrast, wet models are used to describe microswimmers such as synthetic active colloids \cite{Howse_PhysRevLett_2007,VanDerLinden_PhysRevLett_2019,Buttinoni_PhysRevLett_2013}, droplet swimmers \cite{Jin_JPhysCondensMatter_2018,Jin_PNAS_2017,Blois_PhysRevFluids_2019,Maass_AnnRevCondensMattPhys_2016}, and biological microorganisms like bacteria \cite{Elgeti_RepProgPhys_2015,Wolgemuth_CurrBiol_2002,Liu_PhysRevLett_2019}, algae \cite{Rafai_PhysRevLett_2010}, or sperm cells \cite{Elgeti_RepProgPhys_2015}, including their interaction with the surrounding solvent and the corresponding cross interactions among different microswimmers. A particular example of a wet system at larger scales, i.e., beyond the soft matter realm, which microswimmers belong to, can be found in schools of fish \cite{Hemelrijk_InterFoc_2012,Reid_PhysRevE_2012}. To understand the applicability regime of the various active matter models, it is instructive to first discuss the impact of the solvent on active systems: 

\begin{enumerate}[(i)]
    \item \textit{Fluctuations and dissipation:} Active particles are typically orders of magnitude larger than the molecules of the surrounding solvent and are subject to collisions with the latter. This leads to fluctuations in their motion, analogously to Brownian motion of ``passive'' colloids in equilibrium [cf.\ trajectory in Fig.\ \ref{fig:JanusTrajectory} (a)]. The trajectory of an isolated active particle is then typically given by the combination of ballistic motion due to self propulsion and fluctuations due to collisions with the solvent molecules. As can be seen in Fig.\ \ref{fig:JanusTrajectory} (b), the motion of an active particle is not straight because the collisions of the solvent molecules with the active particles feature both a radial and a tangential component. The latter induces a stochastic turn of the particle orientation and, hence, reorientation of the self-propulsion direction, which is called rotational Brownian motion or rotational diffusion. Following the fluctuation-dissipation theorem from statistical mechanics, these fluctuations are necessarily linked to dissipation occurring, e.g., in the form of Stokes drag for spherical particles. For microswimmers, i.e., for (active) particles at the microscale, dissipation normally dominates over inertia. Hence, the motion is overdamped. These effects of the solvent, namely translational diffusion, rotational diffusion, and dissipation, are the only effects of the solvent which are typically taken into account in dry models, such as the active Brownian particle (ABP) model, which we will discuss further below.
    
    \begin{figure}
        \centering
        \includegraphics[width=1.0\linewidth]{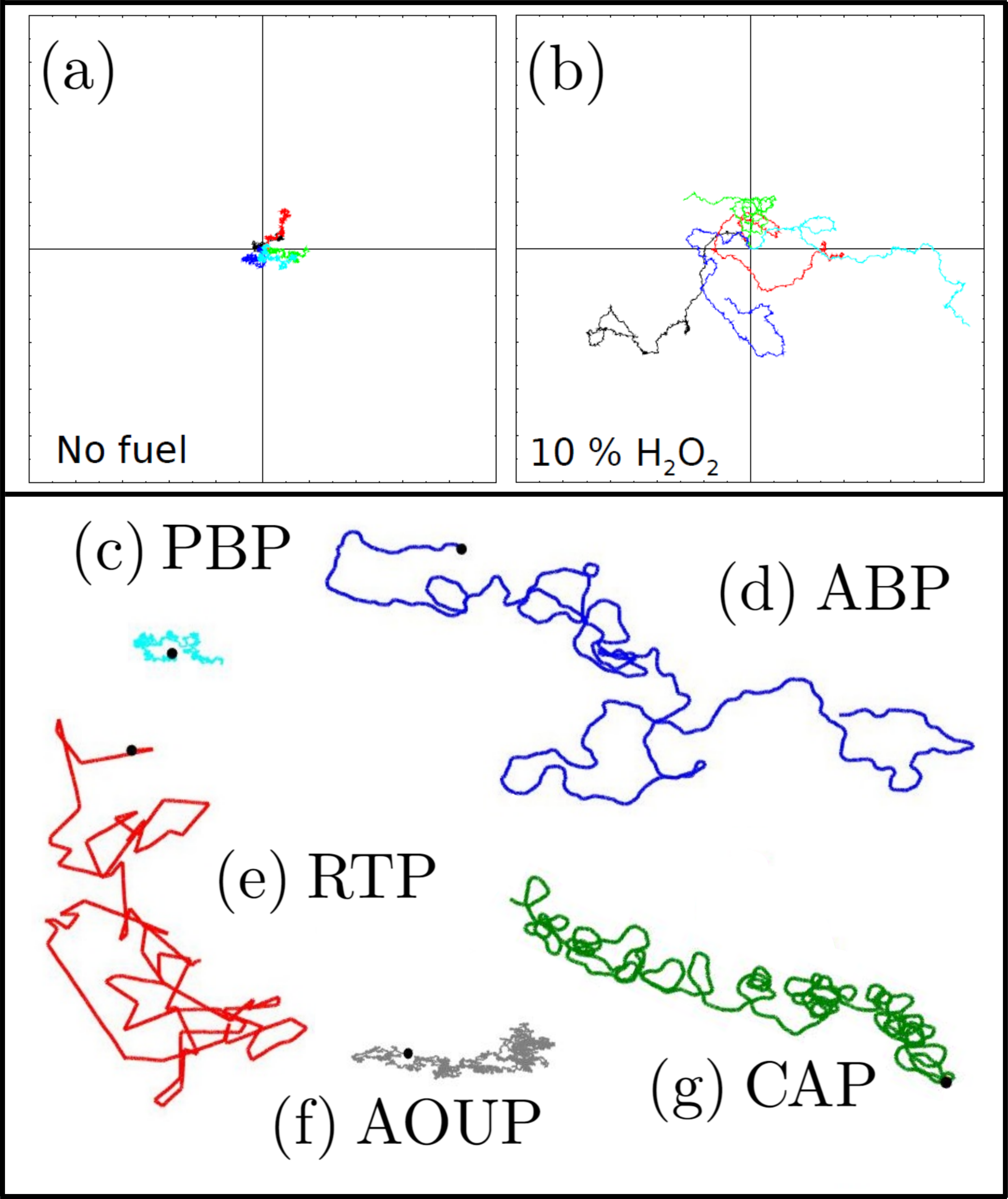}
        \caption{(Top panel) Experimental trajectories of a Janus colloid showing (a) passive and (b) active Brownian motion (kindly provided by J.\ R.\ Howse; see Ref.\ \cite{Howse_PhysRevLett_2007} for experimental details). (Bottom panel) Exemplary trajectories obtained in the overdamped regime from simulations of (c) the passive Brownian particle (PBP) model, (d) the active Brownian particle (ABP) model, (e) the run-and-tumble particle (RTP) model, (f) the active Ornstein-Uhlenbeck particle (AOUP) model, and (g) the chiral active particle (CAP) model. Note that an isolated AOUP is equivalent to an underdamped PBP with $m/\gamma=\tau_{\text{p}}$ (see Sec.\ \ref{sec:DryModels}).}
        \label{fig:JanusTrajectory}
    \end{figure}

    \item \textit{Momentum conservation:} Physically, in the absence of external fields or boundaries, the overall momentum of an active system has to be conserved. For example, when a microorganism or an active Janus colloid moves forward, there is necessarily a counter-propagating solvent flow such that the overall momentum of the active particle and the surrounding solvent is conserved (swimming in vacuum is impossible). Thus, the solvent not only acts as a bath providing fluctuations and drag but also ensures momentum conservation.
    
    \item \textit{Hydrodynamic interactions:} The solvent mediates hydrodynamic interactions among different active particles. These arise because the flow pattern induced by each active particle as a consequence of its swimming acts onto all other particles in the system. These solvent-mediated interactions are often long-ranged. In particular, in the absence of external forces, they often decay as $1/r^2$ for force-dipole swimmers, such as various bacteria or algae (explicit measurements of the flow field exist, e.g., for \textit{E.\ coli} bacteria \cite{Drescher_PNAS_2011}), and as $1/r^3$ for source-dipole swimmers, such as Paramecium \cite{Zhang_EPJST_2015} or (idealized) Janus colloids with a uniform surface mobility \cite{Morrison_1970,interactions_dominate}. However, they can be weakened or decay faster in the presence of a substrate or other boundaries \cite{Lauga_RepProgPhys_2009}.
    
    \item \textit{Hydrodynamic boundary effects:} If the active particles are in contact with boundaries, such as a glass substrate, which is frequently used in experiments with active colloids, or another liquid-solid or liquid-air interface, the solvent can lead to additional interesting effects. An example of these is constituted by osmotic flows at fluid-solid interfaces, such as those induced by auto-phoretic particles \cite{Heidari_Langmuir_2020} or by some modular swimmers involving ion-resins \cite{Niu_SoftMatter_2018,Liebchen_PhysRevE_2018}. At fluid-air interfaces, active particles can cause Marangoni flows \cite{Manjare_JPhysChemC_2015,Fei_CurrOpColIntSci_2017,Marangoni_solution,Dominguez_PRL_2016}, which act on all particles in the system and can elicit interesting collective behaviors \cite{Crowdy_PhysRevFluids_2020,Wittmann_ArXiv_2021,Thutupalli_NewJPhys_2011,Dominguez_SoftMatter_2018}.
\end{enumerate}

\section{Dry active particles: The active Brownian particle model and its alternatives}
\label{sec:DryModels}

\textbf{Active Brownian particle model:} 
\\One of the simplest and most popular models to describe active particles is the active Brownian particle (ABP) model \cite{Review_Romanczuk,TenHagen_CondMatPhys_2009,TenHagen_JPhysCondensMatter_2011,Bechinger_RevModPhys_2016,Inertial_effects_Lowen,models_trajs,ABP_Fodor}, originally introduced to describe the motion of colloidal particles which smoothly change their self-propulsion direction due to rotational diffusion. It treats the solvent as a bath providing only fluctuations and drag without ensuring momentum conservation and, at least in its most commonly used form, without accounting for hydrodynamic interactions among particles. The ABP model does not explicitly describe the mechanism leading to self propulsion either, which arises through the interactions of the active particles with the surrounding solvent (or with a substrate), but simply replaces it with an effective force that drives the particle forward. Microscopically, this is not correct because, as discussed above, microswimmers are force free, but it leads to a simple generic model for the dynamics of active particles, which stays agnostic on many details of the specific underlying realization. In its simplest form, the ABP model in two-dimensional space is defined by the overdamped Langevin equations
\begin{align}
    \frac{\text{d}\vec{r}_i(t)}{\text{d} t} =&~v_0\vec{p}_i(t) - \frac{1}{\gamma}\nabla_{\vec{r}_i} U + \sqrt{2D}\vec{\xi}_i(t), \label{eq:ABP_r} \\ 
    \frac{\text{d}\phi_i(t)}{\text{d} t} =&\sqrt{2D_{\text{R}}}\eta_{i}(t),\label{eq:ABP_phi}
\end{align}
where $\vec{r}_i=(x_i,y_i)$ and $\phi_i$ are the position and the orientation angle of the $i$-th spherical ABP, respectively, $v_0$ is the self-propulsion speed, $\gamma$ is the Stokes drag coefficient, $U=\sum_{i<j}u\left(\left|\vec{r}_i-\vec{r}_{j}\right|\right)$ is the interaction energy with interaction potential $u(r)$, $D$ and $D_{\text{R}}$ are the translational and rotational diffusion coefficients, respectively, and $\vec{\xi}_i(t)=(\xi_{x,i}(t),\xi_{y,i}(t))$ and $\eta_i(t)$ represent Gaussian white noise with unit variance and zero mean. The self-propulsion direction is given by $\vec{p}_i=(\cos{\phi_i},\sin{\phi_i})$.

\begin{figure}
        \centering
        \includegraphics[width=1.0\linewidth]{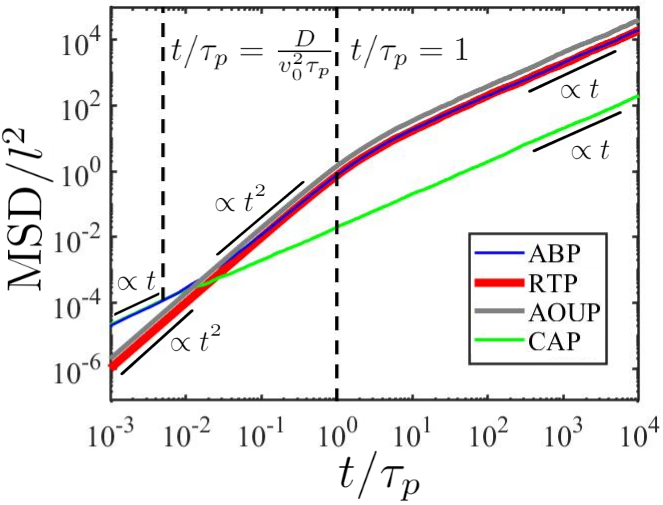}
        \caption{Mean square displacement (MSD) over time in the ABP, RTP, AOUP, and CAP models calculated based on 1000 realizations of the non-dimensionalized equations in Tab.\ \ref{tab:my_label}.
        Simulation parameters: Time step $\Delta t/\tau_{\text{p}}=10^{-3}$, $\text{Pe}=10$ for the ABP and CAP models and $\tilde{\omega}=\pi10^2$ for the CAP model. In the RTP model, the probability that a tumbling event occurs during a certain time step is given by a Poisson distribution with rate $\lambda=\lambda_\text{t}\Delta t=10^{-3}$. The dashed vertical lines are placed at $t/\tau_{\text{p}}=D/(v_0^2\tau_{\text{p}})$ (left) and $t/\tau_{\text{p}}=1$ (right) separating the different regimes exhibited by the particles in each model. The reader is referred to Refs.\ \cite{Chiral_1,Chiral_4} for more details on the MSD of the CAP model.}
        \label{fig:model_trajs_2bis}
\end{figure}

It is instructive to first consider a single ABP, i.e., $U=0$. Different from a passive Brownian particle (PBP) [cf.\ Fig.\ \ref{fig:JanusTrajectory} (c)], the trajectory of an ABP is characterized by an initial period of directed motion followed by a randomization of the self-propulsion direction due to rotational diffusion [cf.\ Fig.\ \ref{fig:JanusTrajectory} (d)]. The initial directed motion persists for a time $\tau_{\text{p}}=1/D_{\text{R}}$ (persistence time) over a distance $l_{\text{p}}=v_0\tau_{\text{p}}$ (persistence length). Accordingly, the average displacement of an ABP reads \cite{TenHagen_JPhysCondensMatter_2011,Bechinger_RevModPhys_2016,Inertial_effects_Lowen}
\begin{equation}\label{eq:average_disp}
    \left\langle \vec{r}_i(t)-\vec{r}_i(0) \right\rangle = l_{\text{p}}\left(1-e^{-t/\tau_{\text{p}}}\right)\vec{p}_i(0).
\end{equation}
Therefore, an ABP moves, on average, over a distance $l_{\text{p}}$ along its initial orientation $\vec{p}_i(0)$ before its orientation is randomized, which rationalizes the term ``persistence lenth''. Additionally, the mean square displacement (MSD) of an ABP (in two-dimensional space) reads \cite{Howse_PhysRevLett_2007,TenHagen_JPhysCondensMatter_2011,Bechinger_RevModPhys_2016,Inertial_effects_Lowen}
\begin{align}\label{eq:MSD}
    \left\langle (\vec{r}_i(t)-\vec{r}_i(0))^2 \right\rangle =& 2l^2_{\text{p}}\left(\frac{t}{\tau_{\text{p}}}-1+e^{-t/\tau_{\text{p}}}\right)\nonumber\\
    &+4Dt,
\end{align}
which provides valuable insight into the different dynamical regimes of the ABP model. Three regimes are observed when expanding Eq.\ (\ref{eq:MSD}) for short, intermediate, and long times. The motion of an ABP is initially diffusive with diffusion coefficient $D$ for $t \ll D/v_0^2$. For $D/v_0^2 \ll t \ll \tau_{\text{p}}$, a ballistic regime which represents directed motion due to the activity of the particle comes about. Finally, for $t \gg \tau_{\text{p}}$, the motion is again diffusive with ``active diffusion coefficient'' $D_{\text{A}}=D+l^2_{\text{p}}/(2\tau_{\text{p}})$. These three regimes are shown in Fig.\ \ref{fig:model_trajs_2bis}.

The relative importance of activity in comparison with diffusion can be characterized by the Péclet number $\text{Pe}=v_0/\sqrt{2DD_{\text{R}}}$ \cite{Bechinger_RevModPhys_2016}. Assuming spherical active particles of radius $R$ and diameter $\sigma=2R$, the Stokes-Einstein relation yields $D=\frac{4}{3}R^2D_{\text{R}}$, and thus the Péclet number can be rewritten in terms of the particle diameter (as customarily done in the literature \cite{Coarsen_law,MIPS_5,Peclet_diam}) as $\text{Pe}=\sqrt{3/2}v_0/(\sigma D_{\text{R}})$ or $\text{Pe}=v_0\sigma/(\sqrt{6}D)$.

\begin{figure}
    \centering
    \includegraphics[width=1.0\linewidth]{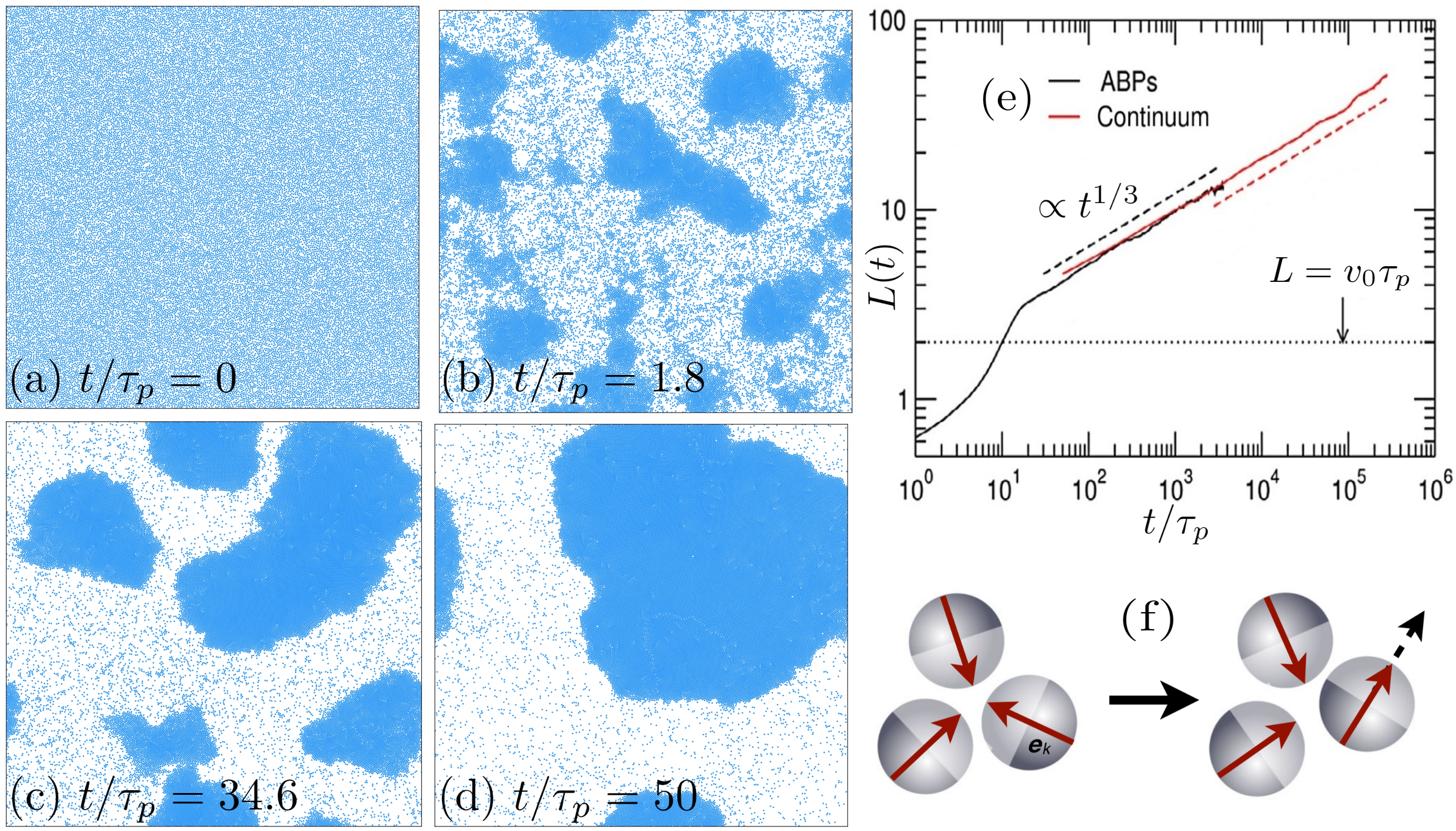}
    \caption{(a)--(d) Evolution of motility-induced phase separation (MIPS) for an ensemble of $N=39200$ ABPs of diameter $\sigma$ with number density $\rho\sigma^3=0.6$, Péclet number $\text{Pe}=200/\sqrt{2}$, and time step $\Delta t/\tau_{\text{p}}=5\times10^{-6}$. The ABPs interact via the purely repulsive Weeks-Chandler-Anderson (WCA) potential \cite{Weeks_JCP_1971} with strength $\epsilon=10\kb T$. (e) Time evolution of the mean cluster size $L(t)$ for the ABP model (adapted from Ref.\ \cite{Coarsen_law} - Published by The Royal Society of Chemistry). (f) Schematic of the mechanism which leads to MIPS (with permission from Ref.\ \cite{Buttinoni_PhysRevLett_2013} - \textcopyright~2013 by the American Physical Society).}
    \label{fig:MIPS}
\end{figure}

For sufficiently dense ensembles of active particles with a sufficiently large P\'eclet number and purely repulsive interactions arising because, e.g., the individual particles cannot overlap, the ABP model predicts a spectacular phenomenon known as motility-induced phase separation (MIPS) \cite{MIPS_1,Buttinoni_PhysRevLett_2013,MIPS_3,MIPS_4,Inertial_effects_Lowen,Caprini_PhysRevLett_2020}. A sequence of snapshots of the state of an ensemble of ABPs which interact via the purely repulsive Weeks-Chandler-Anderson (WCA) potential \cite{Weeks_JCP_1971} and for which MIPS occurs is shown in Fig.\ \ref{fig:MIPS} (a)--(d). Initially, the ensemble is uniformly distributed. For suitable parameters (large Péclet number and high density), the uniform state loses stability and the particles aggregate in small clusters. These clusters grow following the coarsening law shown in Fig.\ \ref{fig:MIPS} (e) until a single macrocluster, which coexists with a low-density active gas, is eventually formed. Overall, while phase separation in equilibrium generally requires inter-particle attractions, active systems can phase separate even in their complete absence \cite{Redner_PhysRevLett_2013,MIPS_4}. The mechanism underlying MIPS is shown in Fig.\ \ref{fig:MIPS} (f): When particles collide, they block each other until their orientations are randomized and they can separate from each other. Broadly, MIPS occurs if the active particles are fast and numerous enough for collisions with existing clusters to occur more often than particles in these clusters leave them due to rotational diffusion.

The ABP model can also be generalized to account for inertial effects \cite{Inertial_effects_Lowen,Mandal_PhysRevLett_2019,Sandoval_PhysRevE_2020,Gutierrez-Martinez_JCP_2020,Petrelli_PhysRevE_2020}, which is used, e.g., to model active granular particles on vibrating plates \cite{Scholz_NatCom_2018,Scholz_NatCom_2018_2}.

\vspace{0.4cm}
\textbf{Alternatives to the ABP model:}
\\Several alternative models have been designed that have a similar scope to that of the ABP model in the sense that they also treat the solvent as a bath which only provides fluctuations and drag rather than accounting for momentum conservation and hydrodynamic interactions.

\vspace{0.2cm}
\textit{Run-and-tumble model:} The run-and-tumble particle (RTP) model \cite{Tailleur_PhysRevLett_2008,Cates_EPL_2013,RTP_3,RTP_4} was originally introduced to describe the characteristic motion patterns of certain bacteria such as \textit{E.\ coli} \cite{Berg_Poisson,Berg_Bacteria,Berg_Book_EcoliInMotion_2004}, but it has now advanced to a standard model for the description of active particles. (In fact, the first theory for MIPS was formulated for RTPs \cite{Tailleur_PhysRevLett_2008} and MIPS has been observed in simulations of this model as well \cite{Cates_EPL_2013}.) In contrast to ABPs, RTPs alternate running periods, during which the self-propulsion direction remains unchanged, with idealized tumbling events, in which the orientation of the particles is randomized [cf.\ Fig.\ \ref{fig:JanusTrajectory} (e)]. The equations of motion for the $i$-th RTP read
\begin{align}
    \frac{\text{d}\vec{r}_i(t)}{\text{d} t} =&~v_0\vec{p}_i(t) - \frac{1}{\gamma}\nabla_{\vec{r}_i} U, \label{eq:RTP} \\ 
    \frac{\text{d}\phi_i(t)}{\text{d} t}=&\sum_{n} \Delta\phi_{n} \delta(t-T_{n}),\label{eq:tumble}
\end{align}
where the parameters of Eq.\ (\ref{eq:RTP}) are defined as in Eq.\ (\ref{eq:ABP_r}). The values of $\Delta\phi_{n}$ are uniformly distributed between $0$ and $2\pi$, with tumbling events taking place at discrete times $T_n$ \cite{RTP_4}. In practice, the times $T_n$ are chosen either randomly with $\langle T_\text{n+1}-T_\text{n} \rangle=\lambda^{-1}_\text{t}$ (and, e.g., tumbling events following a Poisson distribution, which leads to exponentially distributed times between tumbling events, as originally found for \textit{E.\ coli} \cite{Block_JBac_1983}) or equally spaced. In any case, the (mean) tumbling rate $\lambda_{\text{t}}$ is fixed, yielding a persistence time $\tau_{\text{p}}=1/\lambda_{\text{t}}$, which plays the role of the (mean) time between tumbling events. 

Remarkably, the many-particle dynamics following from the RTP and the ABP models turn out to be equivalent at coarse-grained scales if $(d-1)D_{\text{R}}=\lambda_{\text{t}}$, where $d>1$ is the spatial dimension \cite{Cates_EPL_2013,RTP_3}. At the single-particle level, this is also reflected by the MSD in Fig.\ \ref{fig:model_trajs_2bis} showing almost identical time evolution for RTPs and ABPs, with the exception that RTPs always show a ballistic behavior for $t\ll\tau_{\text{p}}$, since tumbling events are statistically unlikely on this timescale and translational diffusion is not considered. The latter can also be taken into account, resulting in the emergence of a diffusive regime for $t\ll D/v^2_0$, as in the ABP model, where $D$ is the translational diffusion coefficient.

\vspace{0.2cm}
\textit{Active Ornstein-Uhlenbeck model:} Another alternative to the ABP model is the active Ornstein-Uhlenbeck particle (AOUP) model \cite{FP_AOUP_1,AOUP_1,AOUP_2,AOUP_3}, which has certain advantages compared with the ABP model regarding the theoretical description of the many-body dynamics of dry active particles. This is due to the fact that the AOUP model avoids the strongly nonlinear dependence of the center-of-mass motion on the particle orientation, which is present in the ABP model [cf.\ Eqs.\ (\ref{eq:ABP_r}) and (\ref{eq:ABP_phi})], by using colored noise to generate self propulsion. The equation of motion for particle $i$ in the AOUP model (in the overdamped regime) reads 
\begin{equation}
    \frac{\text{d}\vec{r}_i(t)}{\text{d} t}=\vec{v}_{0,i}(t)-\frac{1}{\gamma}\nabla_{\vec{r}_i} U,\label{eq:AOUP_r}
\end{equation}
where $\gamma$ is the Stokes drag coefficient and $U$ is the total interaction potential. Whereas the self-propulsion speed $v_0$ remains constant for a single particle in the ABP and RTP model, it evolves with time in the AOUP model according to
\begin{equation}
    \tau_{\text{p}}\frac{\text{d}\vec{v}_{0,i}(t)}{\text{d} t}=-\vec{v}_{0,i}(t)+\sqrt{2D}\vec{\xi}_i(t),\label{eq:AOUP_v}
\end{equation}
where $\tau_{\text{p}}$ is the persistence time and $\vec{\xi}_i(t)$ is Gaussian white noise with unit variance and zero mean. As a result, the velocity components of an isolated AOUP are represented by colored Gaussian noise\footnote{Whereas white noise is delta-correlated in time, the correlation function of colored noise takes finite values for finite time differences.} with correlation function $\left\langle v_{0,i}^{(\alpha)}(0)v_{0,j}^{(\beta)}(t) \right\rangle =\delta_{ij}\delta_{\alpha\beta} (D/\tau_{\text{p}})\e^{-t/\tau_{\text{p}}}$ between components $\alpha$ and $\beta$ of particles $i$ and $j$. Note that a single particle in the AOUP model, i.e., $U=0$, is formally identical to an underdamped passive Brownian particle with $m/\gamma=\tau_{\text{p}}$ [cf.\ Fig.\ \ref{fig:JanusTrajectory} (f)]. Hence, a single AOUP shows a ballistic regime for $t\ll\tau_{\text{p}}$ followed by a diffusive regime for $t\gg\tau_{\text{p}}$, as shown in Fig.\ \ref{fig:model_trajs_2bis}.

Since it involves colored noise, the AOUP model does not permit formulation of an exact Fokker-Planck equation for the corresponding probability distribution. However, it is still possible to derive an approximate Fokker-Planck equation for the many-body dynamics, which does not depend on the particle orientation but only on the particle positions \cite{AOUP_2,AOUP_3,FP_AOUP_1,FP_AOUP_3}. MIPS has also been reported for the AOUP model \cite{AOUP_1,AOUP_3} suggesting that it provides a useful alternative for the description the many-body dynamics of active particles although the single-particle properties significantly differ from those of the ABP and RTP model.

\vspace{0.2cm}
\textit{Chiral particle model:} A further class of models describes chiral active particles (CAPs) \cite{Inertial_effects_Lowen,Liebchen_PhysRevLett_2017_2,Chiral_1,Chiral_2,Chiral_3,Chiral_4,Chiral_5,Chiral_6,Liao_SoftMatter_2018,Bickmann_ArXiv_2020}, which experience an additional effective torque arising from an anisotropy in their shape or propulsion mechanism. For an isolated CAP, this leads to circular trajectories in the limit of zero noise, whereas the orientation angle of the $i$-th CAP in the presence of noise evolves according to
\begin{align}
    \frac{\text{d}\phi_i(t)}{\text{d} t} &= \omega + \sqrt{2D_{\text{R}}}\eta_{i}(t),\label{eq:cap}
\end{align}
where $\omega$ is a constant angular velocity. As in the ABP model, the position $\vec{r}_i$ of the $i$-th CAP generally evolves with time according to Eq.\ (\ref{eq:ABP_r}). An exemplary trajectory of a CAP and the time evolution of its MSD are shown in Figs.\ \ref{fig:JanusTrajectory} (g) and \ref{fig:model_trajs_2bis}, respectively. Examples of circle swimmers include \textit{E.\ coli} bacteria near surfaces and interfaces \cite{Chiral_example_1,Chiral_example_2}, sperm cells \cite{Su_SciRep_2013,Elgeti_BioJ_2010} and artificial microswimmers such as L-shaped particles \cite{Chiral_example_3}, ``spherical-cap particles'' near a substrate \cite{Shelke_Langmuir_2019}, and asymmetric Quincke rollers \cite{Zhang_NatComm_2020}.

An overview of the models introduced thus far is provided in Tab.\ \ref{tab:my_label}. This table is intended as a guide to numerically implementing the previously described models on a single-particle level. To this end, the equations of motion are presented in dimensionless form.

\begin{table*}

\setlength{\arrayrulewidth}{0.5mm}
\setlength{\tabcolsep}{10pt}
\renewcommand{\arraystretch}{2}

\begin{center}
\caption{Dimensionless equations of motion and parameters of a single active particle in the ABP, RTP, AOUP and CAP models. The variables $\vec{\mathfrak{r}}$, $\vec{\mathfrak{v}_0}$, and $\mathfrak{t}$ shown in the equations of the second column have been non-dimensionalized by rescaling the original dimensional variables with respect to the natural time and length scales shown in the fourth column: $\vec{\mathfrak{r}}=\vec{r}/l$, $\vec{\mathfrak{v}_0}=\vec{v}_0\tau_{\text{p}}/l$, $\mathfrak{t}=t/\tau_{\text{p}}$. The dot over these variables denotes the derivative with respect to the dimensionless time $\mathfrak{t}$. Pe is the Péclet number \cite{Bechinger_RevModPhys_2016}.}

\begin{tabular}{ |c|p{3.8cm}|p{2cm}|p{4.5cm}|  }
\hline
\multicolumn{4}{|c|}{\textbf{DRY ACTIVE PARTICLE MODELS FOR A SINGLE PARTICLE}} \\
\hline
\multicolumn{1}{|c|}{\textbf{Model}} & \multicolumn{1}{c|}{\textbf{Equations of motion}} & \multicolumn{1}{c|}{\textbf{Parameters}} & \multicolumn{1}{c|}{\textbf{Natural units}} \\
\hline
\multirow{2}{2em}{ABP} & $\dot{\vec{\mathfrak{r}}}(\mathfrak{t})=\vec{p}(\mathfrak{t})+\text{Pe}^{-1}\vec{\xi}(\mathfrak{t})$ & \multirow{2}{6em}{$\text{Pe}=\frac{v_0}{\sqrt{2DD_{\text{R}}}}$} & Time scale: $\tau_{\text{p}}=D^{-1}_{\text{R}}$ \\ 
& $\dot{\phi}(\mathfrak{t})=\sqrt{2}\eta(\mathfrak{t})$ &  & Length scale: $l=l_{\text{p}}=v_0D^{-1}_{\text{R}}$ \\ 
\hline
\multirow{2}{2em}{RTP} & $\dot{\vec{\mathfrak{r}}}(\mathfrak{t})=\vec{p}(\mathfrak{t})$ & \multirow{2}{1.0\textwidth}{None\footnote{for equally spaced $T_n$ or Poisson-distributed tumbling events as found in \textit{E.\ coli} \cite{Block_JBac_1983} and without translational diffusion; with the latter, the equation of motion for the position reads $\dot{\vec{\mathfrak{r}}}(\mathfrak{t})=\vec{p}(\mathfrak{t})+\text{Pe}^{-1}\vec{\xi}(\mathfrak{t})$ with $\text{Pe}=v_0/\sqrt{2D\lambda_{\text{t}}}$. }}  & Time scale: $\tau_{\text{p}}=\lambda^{-1}_{\text{t}}$ \\ 
& $\Dot{\phi}(\mathfrak{t})=\sum_{n} \Delta\phi_n \delta(\mathfrak{t}-\tilde{T}_n)$ & & Length scale: $l=l_{\text{p}}=v_0\lambda^{-1}_{\text{t}}$ \\
\hline
\multirow{2}{3em}{AOUP} & $\dot{\vec{\mathfrak{r}}}(\mathfrak{t})=\vec{\mathfrak{v}}_0(\mathfrak{t})$ & \multirow{2}{2em}{None} & Time scale: $\tau_{\text{p}}$ \\ 
& $\dot{\vec{\mathfrak{v}}}_0(\mathfrak{t})=-\vec{\mathfrak{v}}_0(\mathfrak{t})+\sqrt{2}\vec{\xi}(\mathfrak{t})$ &  & Length scale: $l=\sqrt{D\tau_{\text{p}}}$ \\
\hline
\multirow{2}{3em}{CAP} & $\dot{\vec{\mathfrak{r}}}(\mathfrak{t})=\vec{p}(\mathfrak{t})+\text{Pe}^{-1}\vec{\xi}(\mathfrak{t})$ & $\text{Pe}=\frac{v_0}{\sqrt{2DD_{\text{R}}}}$ & Time scale: $\tau_{\text{p}}=D^{-1}_{\text{R}}$ \\ 
& $\Dot{\phi}(\mathfrak{t}) = \tilde{\omega} + \sqrt{2}\eta(\mathfrak{t})$ & $\tilde{\omega}=\omega\tau_{\text{p}}$ & Length scale: $l=l_{\text{p}}=v_0D^{-1}_{\text{R}}$ \\
\hline
\end{tabular}
\label{tab:my_label}
\end{center}
\end{table*}

\vspace{0.2cm}
\textit{Monte Carlo simulations:} A final example to describe isotropic dry active particles is based on kinetic Monte Carlo simulations \cite{Monte_Carlo_1,Monte_Carlos_2,Monte_Carlo_3}, where the displacements of the particles are correlated in time. Namely, the displacement during a certain time step is drawn from a Gaussian distribution whose mean equals the displacement in the previous time step.

\begin{figure}
    \centering
    \includegraphics[width=0.97\linewidth]{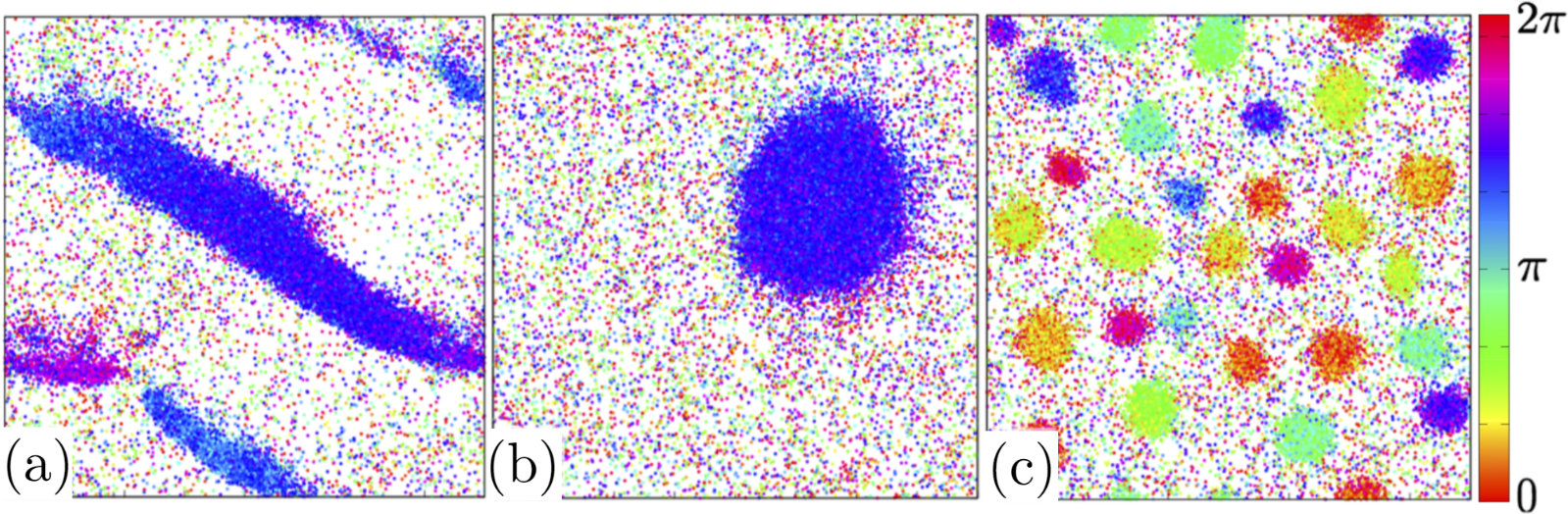}
    \caption{Simulation snapshots of active particles with alignment interactions [cf.\ Eqs.\ (\ref{eq:Vicsek_r}) and (\ref{eq:Vicsek_phi})] for $D=0$ and (a) $\omega=0$ (smooth variant of the Vicsek model) and (b), (c) $\omega>0$ (chiral active particles). Colors represent particle orientation angles such that equally colored particles are aligned or phase-synchronized (with permission from Ref.\ \cite{Liebchen_PhysRevLett_2017_2} - \textcopyright~2017 by the American Physical Society).}
    \label{fig:Vicsek}
\end{figure}

\vspace{0.2cm}
\textbf{Models with explicit alignment interactions:} 
Thus far, we have focused our discussion on isotropic active particles, i.e., on particles without explicit alignment interactions. The most popular model for describing (dry) active particles with (polar) alignment interactions is the Vicsek model \cite{Vicsek_PhysRevLett_1995,Phys_Vic_Mod}, which accounts for self-propelled particles (``birds'') that align their orientation with that of their neighbors. A generalized continuous-time variant of the Vicsek model comprising CAPs with alignment interactions can be defined by 
\begin{align}
    \frac{\text{d}\vec{r}_i(t)}{\text{d} t} =&~ v_0\vec{p}_i(t) + \sqrt{2D}\vec \xi_i(t),\label{eq:Vicsek_r}\\ 
    \frac{\text{d}\phi_i(t)}{\text{d} t} =&~\omega + \frac{K}{\pi R^2_0} \sum_{j\in S_{R_0}^{(i)}} \sin(\phi_j-\phi_i)\nonumber\\
    &+ \sqrt{2D_{\text{R}}}\eta_i(t),\label{eq:Vicsek_phi}
\end{align}
where $\omega$ is an angular velocity, $K$ is the strength of the alignment interactions (for K=0 this model reduces to the CAP model) and the sum is calculated over all particles within a circle $S_{R_0}^{(i)}$ of radius $R_0$ centered at the position of particle $i$ \cite{Vicsek_mod_1,Farrell_PhysRevLett_2012,Vicsek_mod_3,Liebchen_PhysRevLett_2017_2}. The hallmark of this model is that particles tend to follow the orientation of their neighbors, which can induce polar order, e.g., in the form of the traveling bands shown in Fig.\ \ref{fig:Vicsek} (a) for $\omega=0$. When considering CAPs with polar interactions (Eq.\ (\ref{eq:Vicsek_phi}) with $\omega>0$), one finds two remarkable phenomena: The formation of rotating macrodroplets with late-time sizes comparable to the system size, which is indicative of phase separation, at low angular velocity [cf.\ Fig.\ \ref{fig:Vicsek} (b)]; and a pattern of rotating microflocks exhibiting phase synchronization and a self-selected length scale at high angular velocity [cf.\ Fig.\ \ref{fig:Vicsek} (c)].

Another class of models with alignment interactions accounts for nematic alignment interactions \cite{Marchetti_RevModPhys_2013,Nematic_1,Nematic_2,Nematic_3}, which arise in systems of head-tail symmetric particles, such as (self-propelled) rods \cite{Baer_AnnuRevCondensMatterPhys_2020} featuring apolar interactions.

\vspace{0.2cm}
\textbf{Applicability regime of dry active particle models:}
\\The ABP model and its alternatives are commonly used to perform particle-based simulations of active particles and also as a starting point for the formulation of continuum theories, as we shall discuss hereunder. These models have proven useful when applied to, e.g., the following problems concerning active particles:

\begin{enumerate}[(i)]
    \item When we are concerned with single active particle flow fields, the ABP model has been very successful, e.g., to predict correlation functions in close agreement with experiments of Janus colloids \cite{Kurzthaler_PhysRevLett_2018}.
    
    \item When hydrodynamic interactions play a minor role such as for certain active colloids, their many-body behavior is reasonably well described by ABPs \cite{Buttinoni_PhysRevLett_2013}. Similarly, when hydrodynamic interactions are dominated by other interactions such as, e.g., phoretic interactions of autophoretic colloids with a near-uniform surface mobility, the ABP model serves as a useful starting point for the derivation of simple models with effective phoretic pair interactions \cite{interactions_dominate}.

    \item When a solvent is absent but fluctuations are still relevant as, e.g., for granular particles on vibrating plates, where quasi-deterministic chaos arises and leads to effective randomness, which can be described as Brownian noise, the ABP model can be used as a numerical model \cite{Lanoiselee_PhysRevE_2018,Walsh_SoftMatter_2017}.

    \item The ABP model is also useful for fundamental theoretical explorations, e.g., when we are more interested in the fundamental consequences of activity on the collective behavior of active particles rather than in the specific link to experimental realizations. 

\end{enumerate}

\vspace{0.2cm}
\textbf{Advantages and limitations:} 
\\Compared with most ``wet'' models, a key advantage of the ABP model and its alternatives is their simplicity from both a conceptual and a computational viewpoint. In particular, these models allow one to simulate very large ensembles of active particles (state-of-the-art simulations often use $10^5-10^7$ particles \cite{Caporusso_PhysRevLett_2020,Mandal_PhysRevLett_2019,Stenhammar_PhysRevLett_2013,Coarsen_law,Digregorio_PhysRevLett_2018,Redner_PhysRevLett_2013}). One key limitation of these models regarding the description of soft active matter systems is that they do not account for momentum conservation and often not for hydrodynamic interactions either. This can be particularly relevant for the description of the collective behavior or for describing single microswimmers near walls. The ABP model is popular when simulating the collective behavior of autophoretic active colloids as well. Here, beside hydrodynamic interactions, also phoretic interactions can play a crucial role and are also neglected by the standard ABP model \cite{interactions_dominate}, which can however be extended to take them into account \cite{Liebchen_PhysRevLett_2017,Phoretic_1,Phoretic_2,Phoretic_3,Phoretic_4,Phoretic_5,Phoretic_6}.

\section{Continuum theories for dry active matter}

To understand the collective behavior of (dry) active particles, one often uses continuum models, which can be used for a purely theoretical analysis or a numerical analysis based on continuum simulations. In general, one can distinguish between (i) phenomenological and (ii) microscopic theories. 

\begin{itemize}
    \item[(i)] \textbf{Phenomenological theories:} This class is often based on an identification of the relevant ``slow variables'' (e.g., the density field $\rho(\vec{r},t)$ in the case of isotropic active systems with particle number conservation or the density field and polarization density for polar active systems with polar alignment interactions) and on writing down all terms which are allowed by symmetry and conservation laws up to a certain order. Accordingly, these theories are sometimes called Landau theories. A key advantage of phenomenological theories is that they predict the structure of the field equations essentially based on symmetry, conservation laws, and dimensionality of the system without requiring any reference to the details of the underlying particle system (such as the precise form of the interactions). Thus, these field theories are sometimes called ``generic'' in this sense and can even be formulated (and numerically solved) if no underlying particle-based model is known. However, phenomenological field theories do not provide information about the values of the coefficients. Thus, one often treats all occurring coefficients as independent parameters and studies the phenomenology of the field equations as a function of all these parameters. A related important drawback of this approach is that it then remains unclear if there is an underlying particle-based model or realization which leads to the corresponding parameter values. A specific example of a phenomenological theory is discussed further below.

    \item[(ii)] \textbf{Microscopic theories:} In contrast to phenomenological theories, microscopic theories involve a systematic derivation of the field equations typically from the underlying equations of motion for the individual active particles. This approach yields equations of motion for the relevant fields, which directly follow from the underlying particle-based model. Thus, in contrast to the former class of theories, one advantage of this second approach is that one obtains, in addition to the structure of equations, an explicit link between the coefficients of the particle-based model and the continuum theory. This typically leads to a (much) smaller number of independent parameters than one would obtain from phenomenological approaches. Another advantage is that, following the microscopic approach, terms which are allowed by symmetry cannot be missed, which has happened for various standard models of active matter in the past when following the phenomenological approach.
\end{itemize}

In the following, we will illustrate both approaches based on specific examples for isotropic and polar active systems.

\vspace{0.6cm}
\textbf{Example: Phenomenological theory for isotropic active matter}
\\Collective phenomena of isotropic active matter, such as phase separation, can be described, e.g., by the phenomenological active model B+, which is based on the common model B that describes phase separation in equilibrium systems \cite{Hohenberg_RevModPhys_1977}. Here, the density field $\rho(\vec{r},t)$ is assumed to be the only slow variable of the system and the order parameter $\phi$ is related to it by the linear transformation $\phi=(2\rho-\rho_{\text{H}}-\rho_{\text{L}})/(\rho_{\text{H}}-\rho_{\text{L}})$, where $\rho_{\text{H}}$ and $\rho_{\text{L}}$ denote the density at the low-density and the high-density critical point, respectively \cite{Shaebani_NatRevPhys_2020,Wittkowski_NatComm_2014}. The active model B+ is given by the equations
\begin{align}
    \frac{\partial \phi}{\partial t}=&-\nabla\cdot\left[-M\nabla\left(\frac{\delta\mathcal{F}}{\delta\phi}+\lambda|\nabla\phi|^2\right)\right.\nonumber\\
    &\left.+ \zeta M\left(\nabla^2\phi\right)\nabla\phi + \sqrt{2D}\vec{\Lambda}\right], \label{eq:BPlus_Continuity}\\
    \mathcal{F}[ \phi ] =& \int\text{d}^3r\,\left[ \frac{a}{2}\phi^2+\frac{b}{4}\phi^4+\frac{K}{2}|\nabla\phi|^2 \right]. \label{eq:BPlus_FreeEnergy}
\end{align}

\begin{figure}
    \centering
    \includegraphics[width=1.0\linewidth]{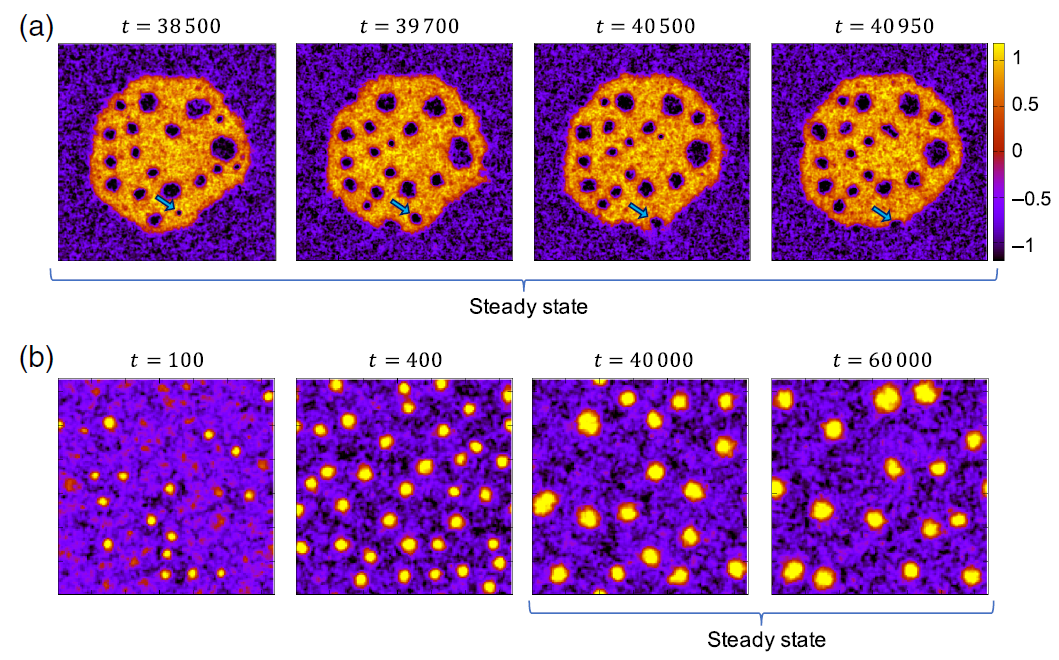}
    \caption{Numerical results of the active model B+: (a) Coexistence of a liquid phase (yellow to red colors) comprising continuously created vapor bubbles and a gas phase (black and purple). (b) Phase separation into a dense (yellow to red colors) and a dilute phase (black and purple). Dense clusters stabilize at a certain cluster size in the steady state (taken from Ref.\ \cite{Tjhung_PhysRevX_2018}).}
    \label{fig:BPlus_Phases}
\end{figure}

Here, the free-energy functional $\mathcal{F}$ is approximated up to the order $\phi^4$ and up to square-gradient terms \cite{Tjhung_PhysRevX_2018}. Equation (\ref{eq:BPlus_Continuity}) has the form of a continuity equation, and hence it ensures particle number conservation, whereas reaction terms are not allowed in Eq.\ (\ref{eq:BPlus_Continuity}). The order parameter $\phi$ is subject to a Gaussian white noise field $\vec{\Lambda}(\vec{r},t)$ with zero mean and unit variance. The diffusion coefficient is denoted by $D$ and $M$ is the mobility of the active particles. For active particles, the time-reversal symmetry (TRS) is broken locally. This fact is included in the active model B+ by the additional terms proportional to $\lambda$ and $\zeta$. The active model B+ describes the phase separation behavior of isotropic active matter and predicts two types of patterns: The first one is characterized by phase separation into a dense and a dilute phase and the additional occurrence of vapor bubbles inside the dense phase, which are continuously created and move to the surface of the dense phase [cf.\ Fig.\ \ref{fig:BPlus_Phases} (a)]. The second pattern is characterized by the emergence of dense clusters that do not grow beyond a certain characteristic size [cf.\ Fig.\ \ref{fig:BPlus_Phases} (b)]. The coefficients $a,b,K,\lambda,\zeta$ are not known in this phenomenological approach and are treated as parameters of the model. Thus, there is no obvious connection to particle-based models such as the ABP model, whereas in microscopic theories all parameters are directly related to the underlying particle-based model, as will be discussed next.

\begin{figure}
    \centering
    \includegraphics[width=1.0\linewidth]{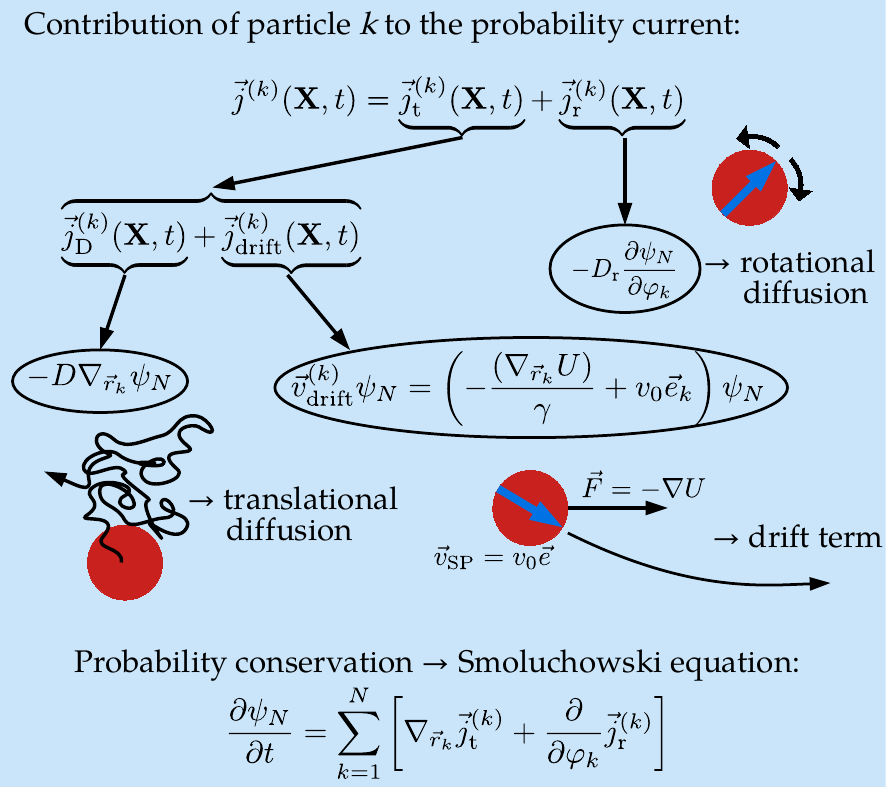}
    \caption{Interpretation of the Smoluchowski equation [cf.\ Eq.\ (\ref{eq:SmoluchowskiEqABPs})] for an ABP as probability conservation law. The contribution of each particle to the probability current can be decomposed into a translational and a rotational current. The former includes translational diffusion and a drift term due to the interaction potential as well as the self-propulsion velocity, whereas the latter considers rotational diffusion. The Smoluchowski equation can then be interpreted as a continuity equation ensuring probability conservation.}
    \label{fig:SmoluchowskiInterpretation}
\end{figure}

\vspace{0.1cm}
\textbf{Example: Microscopic theories for isotropic active matter}
\\There are several approaches to developing microscopic theories. To exemplify one of them, we consider a system of $N$ active particles without alignment interactions. Then, we write down the Smoluchowski equation for the $N$-particle probability density and integrate out variables to obtain the one-particle density field. This approach has been used, for instance, to formulate a microscopic theory of MIPS in overdamped ABPs with positions $\vec{r}_i$ and orientations $\varphi_i$ \cite{Bialke_EPL_2013}. Let $\mathbf{X}=\{\vec{r}_1,...,\vec{r}_N,\varphi_1,...,\varphi_N\}$ denote the state of the $N$-particle system. The corresponding Smoluchowski equation \cite{Risken_Book_TheFokker-PlanckEquation_1984} for the joint probability distribution $\psi_N(\mathbf{X},t)$ reads
\begin{align}
    \frac{\partial \psi_N}{\partial t}=&\Sum{k=1}{N}\nabla_{\vec{r}_k}\cdot\left[\frac{\left(\nabla_{\vec{r}_k}U\right)}{\gamma}-v_0\vec{p}_k+D\nabla_{\vec{r}_k}\right]\psi_N\nonumber\\
    &+D_{\text{R}}\Sum{k=1}{N}\frac{\partial^2\psi_N}{\partial \varphi_k^2},\label{eq:SmoluchowskiEqABPs}
\end{align}
with $U=\sum_{k<k'}u\left(\left|\vec{r}_k-\vec{r}_{k'}\right|\right)$, interaction potential $u(r)$, and self propulsion along $\vec{p}_k$ with speed $v_0$. The Smoluchowski equation ensures probability conservation and its physical interpretation is illustrated in Fig.\ \ref{fig:SmoluchowskiInterpretation}. Starting from Eq.\ (\ref{eq:SmoluchowskiEqABPs}), one usually derives an equation of motion for the one-particle probability distribution $\psi_1(\vec{r}_1,\varphi_1,t)$ by integration. Due to the pair interactions, the resulting equation still contains terms which include the two-particle probability distribution. Similarly, one can derive an equation for the two-particle probability distribution, which then includes the three-particle distributions, and so on, leading to a hierarchy of coupled differential equations that have to be closed by a suitable closure scheme \cite{Book_TheoryOfSimpleLiquids_Hansen_2006,Bertin_JPhysAMathTheor_2009}. Afterwards, an equation of motion for the particle density $\rho(\vec{r},t)$ can be derived by integrating over the orientation $\varphi$, which typically couples again to higher moments and leads to a second hierarchy of equations, which has again to be closed using a suitable closure scheme. To study phase separation, one possible approximation to avoid the first type of hierarchy is to assume that the density varies slowly in space such that the local density is constant within the range of the interaction potential resulting in an effective self-propulsion speed $v(\rho)=v_0-\zeta\rho$ with constant $\zeta$. This density-dependent self-propulsion speed effectively accounts, to some extend, for the net effect of the repulsive interactions, namely the slowdown of active particles in regions of high density. The result of this microscopic approach fits well to computer simulations of ABPs and predicts MIPS in overdamped ABPs \cite{Bialke_EPL_2013}.

An alternative approach, sometimes called the ``Dean approach'' \cite{Dean_JPhysA_1996}, is based on an explicit coarse-graining of the Langevin equations for the individual particles. This approach has been applied in several works, e.g., to describe MIPS in systems of RTPs \cite{Tailleur_PhysRevLett_2008}, pattern formation in self-propelled particles with alignment interactions \cite{Farrell_PhysRevLett_2012}, collective phenomena in systems of CAPs \cite{Liebchen_PhysRevLett_2017_2}, pattern formation in systems of phoretically interacting active colloids \cite{Liebchen_PhysRevLett_2017}, or active systems showing nematic order \cite{Bertin_NewJPhys_2013}. Here, one uses Itô calculus \cite{Dean_JPhysA_1996,Gardiner_Book_StochasticMethods_2009,Chibbaro_Book_StochasticMethodsInFluidMechanics_2014} to deduce a stochastic differential equation, which involves multiplicative noise, for the (fluctuating) combined probability $f(\vec{r},\varphi,t)=\sum_{i=1}^{N}\delta(\vec{r}_i(t)-\vec{r})\delta(\varphi_i-\varphi)$ to find one particle with orientation $\varphi$ at position $\vec{r}$ at time $t$. To derive the one-particle density field $\rho(\vec{r},t)$, one can then, for example, choose to neglect the multiplicative noise term (mean field) and derive a hierarchy of equations in a similar way to the Smoluchowski approach.

\vspace{0.1cm}
\textbf{Example: Microscopic theories for polar active matter}
\\The aforementioned continuum theories for dry active matter were focused on isotropic active matter that can be described by only considering the density field. However, if the particles feature alignment interactions such as in the Vicsek model, polar order can arise. Thus, describing these systems additionally requires the consideration of the mean local orientation of the particles by means of a polarization density $\vec{p}(\vec{r},t)$. Corresponding theories for the density field and the polarization density can be derived based on the Smoluchowski approach or the Dean approach. Another approach, which is aimed at describing the collective behavior of the Vicsek model (which is discrete in time in its original formulation) and is given by Ref.\ \cite{Ihle_PhysRevE_2011}, is based on the Liouville equation for the $N$-particle probability density $\psi_N(\mathbf{X},t)$ and applied to the well-known Vicsek model \cite{Vicsek_PhysRevLett_1995}. Within this model, the particles only interact during a collision event by aligning their orientation to that of their next neighbors and the orientation is subject to Gaussian white noise. Under the assumption that the particles are uncorrelated prior to a collision, the $N$-particle density is written as a product of one-particle densities, which is a good approximation if the noise strength is large and if the mean-free path between two collisions is larger than the interaction radius. Then, the one-particle probability distribution is obtained by integration. However, the solution contains complicated collision integrals that are approximated using the Chapman-Enskog expansion \cite{Ihle_PCCP_2009}, which takes the stationary state as a reference and expands around it in powers of the gradients. Finally, this leads to a set of two coupled differential equations for $\rho$ and $\vec{p}$. This set of equations is similar to that of the phenomenological Toner-Tu model \cite{Toner_PhysRevE_1998} except for additional gradient terms, which occur only in the microscopic approach.

Independently of whether a theory is phenomenological or microscopic, the relevant field equations can then be studied based on various analytical and numerical techniques ranging from perturbation theories, linear stability analyses, or dynamical renormalization group calculations in the presence of additional noise terms to explicit numerical solutions based on, e.g., finite difference, finite volume, or finite element methods.

\section{Hydrodynamics of microswimming: Low Reynolds number and Stokes flow}

The ABP model and its alternatives do not resolve the self-propulsion mechanism, but instead involve an effective force to phenomenologically model the resulting directed motion. To understand and describe the self-propulsion mechanism of a microswimmer, one has to explicitly model the flow field produced by the microswimmer and its interaction with the body of the swimmer.
 
\vspace{0.2cm}
\textbf{Microhydrodynamics:} Let us now briefly discuss the basic equations which are involved in the modeling of a single microswimmer. While swimming at the macroscale involves inertia and leads to flow fields which are described by the Navier-Stokes equation, microswimmers have to employ swimming mechanisms which work even in the absence of inertia since, at the microscale, viscous effects dominate over inertial effects. This is quantified by the Reynolds number, which measures the relative importance of inertial and viscous forces and is given by $\mathrm{Re}=(\rho L v)/\eta$, where the numerator represents the product of the fluid density, a characteristic length scale, and a typical flow speed, whereas the denominator contains the solvent viscosity. For microswimmers, $\mathrm{Re}\ll 1$: For \textit{E.\ coli} bacteria in water, for example, we have $L\sim 3\,\mu\text{m},~v\sim 30\,\mu\text{m/s},~\eta = 0.001\,\text{Pa s}$, and $\rho=1\,\text{g/cm}^3$ \cite{Bechinger_RevModPhys_2016}. Thus, $\text{Re}\sim 10^{-5}-10^{-4} \ll 1$ and inertial effects can safely be neglected. For comparison, phenomena occurring at high Reynolds numbers, such as turbulence, often occur at $\text{Re}\sim 10^3-10^4$ \cite{book_Hydrodynamik_Wolschin_2016}.

\begin{figure}
    \centering
    \includegraphics[width=1.0\linewidth]{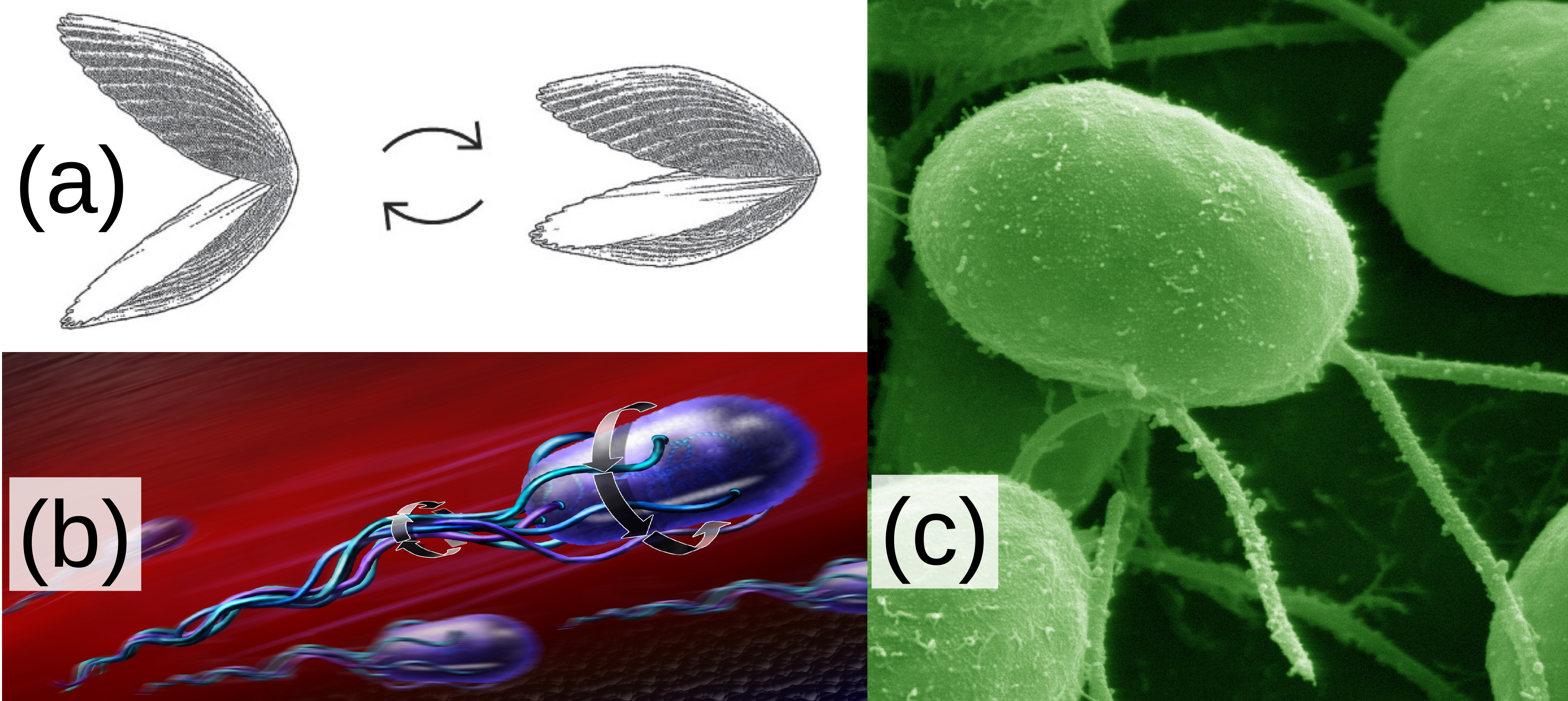}
    \caption{(a) Motion of a scallop. By quickly closing and slowly opening the two shells, the scallop produces a net flow and starts moving. At low Reynolds number, the net displacement is zero for this reciprocal motion (taken from Ref.\ \cite{Qiu_NatComm_2014}). (b) Schematic of the non-reciprocal motion of an \textit{E.\ coli} bacterium (taken from Ref.\ \cite{escherichia_coli_fig}). (c) Electron microscope image of \textit{Chlamydomonas reinhardtii} algae showing the flagella producing self propulsion by non-reciprocal motion (taken from Ref.\ \cite{chlamydomonas_reinhardtii_fig}).}
    \label{fig:Scallop}
\end{figure}

At low Reynolds number, the Navier-Stokes equation reduces to the Stokes equation, which describes ``creeping flow'' and reads 
\begin{equation}
    \eta \nabla^{2} \vec{u}-\nabla p + \vec{f}=0,
    \label{eq:StokesEq}
\end{equation}
where $\vec{u}(\vec{r},t)$ and $p(\vec{r},t)$ are the solvent velocity field and the pressure field, respectively, and $\vec{f}(\vec{r},t)$ is the force density representing the forces exerted by the microswimmers on the solvent. The Stokes equation is typically complemented by the incompressibility condition $\nabla \cdot \vec{u}=0$ leading to a complete set of equations to determine $\vec{u}(\vec{r},t)$ and $p(\vec{r},t)$ for a given $\vec{f}(\vec{r},t)$ and given boundary conditions. Notably, the Stokes equation does not contain any time derivatives, and therefore, the solvent responds instantaneously to the applied forces (no motion would take place once the forcing term is switched off), which reflects the absence of inertia. Accordingly, the swimming mechanism of scallops, which move by periodically opening and closing their shells, would not work at low Reynolds number [cf.\ Fig.\ \ref{fig:Scallop} (a)]. Likewise, any other mechanism based on reciprocal motions would not lead to directed motion. This is Purcell's scallop theorem \cite{Purcell_AmericanJPhys_1977}. 

The general procedure to model microswimmers which move by body-shape deformations (or squirmers) at low Reynolds numbers consists in solving the Stokes equation with appropriate boundary conditions for the solvent velocity field $\vec{u}$ on the surface of the microswimmers. This yields the solvent velocity field $\vec{u}$, from which the stress tensor $\mathbf{\sigma}=\eta\left(\nabla\otimes\vec{u}+(\nabla\otimes\vec{u})^{\text{T}}\right)$ can be obtained. The latter then allows one to calculate the total force $\vec{F}=\int_{S}\text{d}S\,\mathbf{\sigma}(\vec{r},t)\hat{n}$ and the torque $\vec{T}=\int_{S}\text{d}S\,\vec{r}\times(\mathbf{\sigma}(\vec{r},t)\hat{n})$ which act on the microswimmer, where $S$ and $\text{d}S$ denote the surface of the microswimmer and a differential element of it, respectively. Then, for a solid particle, the rigidity condition
\begin{equation}
    \vec{u}(\vec{r})=\vec{v}+\vec{\omega}\times\vec{r},\quad\vec{r}\in S
    \label{eq:StokesBoundary}
\end{equation}
is typically assumed to apply at the surface $S$ of the particle and links the particle velocity $\vec{v}$ and angular velocity $\vec{\omega}$ to $\vec{F}$ and $\vec{T}$. Finally, the torque-free ($\vec{T}=0$) and force-free ($\vec{F}=0$) conditions allow one to solve for $\vec{v}$ and $\vec{\omega}$ \cite{Bechinger_RevModPhys_2016,Zoettl_JPhysCondensMatter_2016}. Since microswimmers often deform in a cyclic way, the net displacement during one cycle of period $T$ is given by $\int_{0}^{T}\text{d}t\,\vec{v}(t)$, which is zero for reciprocal movement in the regime of low Reynolds numbers \cite{Lauga_RepProgPhys_2009}. Thus, non-reciprocal body-shape deformations are required to produce directed motion. Two examples of biological microswimmers that self propel by non-reciprocal motion are demonstrated in Fig.\ \ref{fig:Scallop} (b) and (c). A minimal microswimmer model can be constructed, e.g., based on three spheres connected by two arms, which periodically change their length (three-sphere swimmer) \cite{Golestanian_PhysRevE_2008,Alouges_EurPhysJE_2009,Nasouri_PhysRevFluids_2019,Daddi-Moussa-Ider_JCP_2018,Daddi-Moussa-Ider_JPhysCondensMatter_2018} or based on two spheres which can contract or expand radially and are connected by an elastic arm \cite{Alouges_EurPhysJE_2009,Wang_JMathBiol_2018,Silverberg_BiopinspirBiomim_2020}.




\section{Modeling hydrodynamics at the many-particle level}

In ensembles of microswimmers, each of them generates a specific flow pattern which typically decays slowly in space and leads to long-ranged hydrodynamic cross interactions among different microswimmers as well as to hydrodynamic (self) interactions with walls and interfaces. These hydrodynamic interactions are typically not included in models of dry active matter such as the ABP model and its alternatives. One way of simulating several interacting microswimmers is to explicitly model the detailed self-propulsion mechanism of each microswimmer, i.e., to alternately solve the Stokes equation with the microswimmer-solvent boundary conditions for all swimmers simultaneously and to propagate the swimmers based on the force- and torque-free conditions. While such an approach is conceptually relatively simple and accurate in principle, it creates a huge numerical effort and typically becomes unfeasible even for moderately large microswimmer ensembles. In the following, we briefly discuss some alternative approaches, which allow for more efficient numerical descriptions of microswimmer ensembles.


\vspace{0.2cm}
\textbf{Minimal models and hydrodynamic far-field interactions:} 
\\To model the dynamics of large microswimmer ensembles, an explicit modeling of the solvent flow including the detailed particle-solvent boundary conditions occurring in real microswimmers is often numerically so demanding that very large system sizes remain unreachable. Therefore, one often looks for a compromise between the ABP model, which neglects hydrodynamic interactions and momentum conservation altogether, and an explicit modeling of the self-propulsion mechanism of all interacting microswimmers in a given ensemble. One common approach involves formulating hydrodynamically consistent minimal models for the collective behavior of microswimmers, where one does not explicitly describe the self-propulsion mechanism of each microswimmer but replaces each microswimmer with a simpler representative that creates a similar (far-field) flow pattern. To this end, one uses a multipole expansion of the flow field (similar to that used, e.g., in electrodynamics) \cite{Yeomans_RivNuovoCim_2017,Graham_Book_Microhydrodynamics_2018,Kim_Book_Microhydrodynamics_1991,Winkler_InBook_HydrodynamicsInMotileActiveMatter_2018} and only considers the leading-order terms. In the simplest case, these are the so-called ``singularity solutions'' of the Stokes equation (e.g., the flow field of a force dipole), which are then used to replace the flow field created by each microswimmer and are equivalent to the far-field flow pattern generated by the actual microswimmer to be modeled. For example, it is well known that \textit{E.\ coli} bacteria produce essentially the same far-field flow pattern as a force dipole (pusher) \cite{Drescher_PNAS_2011} and \textit{Chlamydomonas} algae produce a far-field flow pattern which can be represented by the flow field produced by an oscillatory force dipole \cite{Guasto_PhysRevLett_2010}. Let us briefly discuss three common singularity solutions of the Stokes equation:
\begin{itemize}
    \item[(i)] \textit{Point force (``Stokeslet''):} The flow generated by a point force $\vec{f}_{\text{p}}=f\hat{e}\delta(\vec{r}-\vec{r}_0)$ placed at position $\vec{r}_0$ and pointing along the direction $\hat{e}$ is similar to the far-field flow of a particle that is driven by an external force \cite{Zoettl_JPhysCondensMatter_2016,Graham_Book_Microhydrodynamics_2018}. By setting $\vec{f}=\vec{f}_{\text{p}}$ in the Stokes equation [cf.\ Eq.\ (\ref{eq:StokesEq})], the resulting velocity field reads
    \begin{equation}
        \vec{u}_{\text{PF}}(\vec{r})=\frac{f}{8\pi\eta r}\left[\hat{e}+\left(\hat{r}\cdot\hat{e}\right)\hat{r}\right],
        \label{eq:PointForce}
    \end{equation}
    where $r=|\vec{r}|$, $\hat{r}=\vec{r}/r$, and $\eta$ denotes the viscosity of the solvent. The velocity field is shown in Fig.\ \ref{fig:StokesSolutionDipoles} (a). Since microswimmers are force free (momentum conservation), the Stokeslet solution alone is unsuitable to represent them.
    
    \begin{figure}
        \centering
        \includegraphics[width=1.0\linewidth]{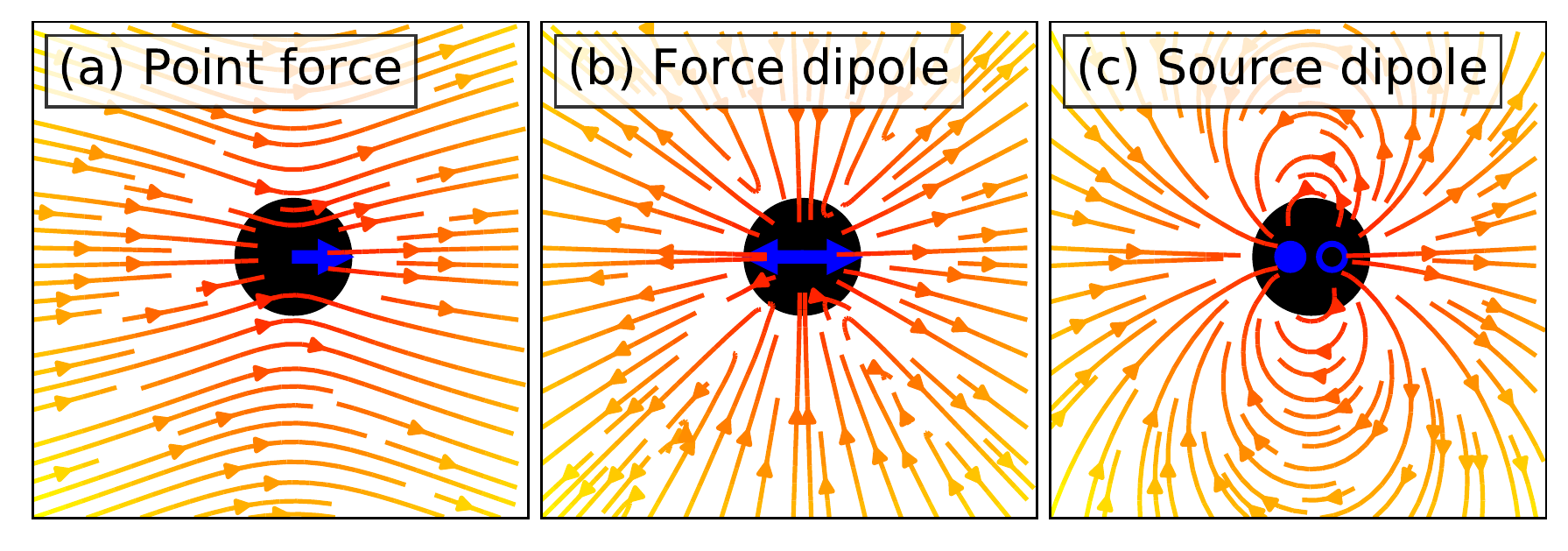}
        \caption{Illustration of the velocity field $\vec{u}(\vec{r})$ of (a) a point force, (b) a force dipole, and (c) a source dipole.}
        \label{fig:StokesSolutionDipoles}
    \end{figure}
    
    \item[(ii)] \textit{Force dipole:} The far-field solution of the Stokes equation in the presence of two point forces $\vec{f}_+=f\hat{e}\delta(\vec{r}-\vec{r}_0-(l/2)\hat{e})$ and $\vec{f}_-=-f\hat{e}\delta(\vec{r}-\vec{r}_0+(l/2)\hat{e})$, which are separated by a distance $l$, reads 
    \begin{equation}
        \vec{u}_{\text{FD}}(\vec{r})=\frac{fl}{8\pi\eta r^2}\left[3\left(\hat{e}\cdot\hat{r}\right)^2-1\right]\hat{r}
        \label{eq:ForceDipole}
    \end{equation}
    in the limit $l\rightarrow 0$ or at distances $r\gg l$ and it is represented in Fig.\ \ref{fig:StokesSolutionDipoles} (b) for $f>0$ \cite{Zoettl_JPhysCondensMatter_2016,Graham_Book_Microhydrodynamics_2018}. These force dipoles push fluid molecules in the forward and backward directions. Hence, microswimmers that show this kind of far-field flow pattern are called ``pushers'', whereas the case of $f<0$, where all flow field lines are reverted, corresponds to a ``puller''. 
    
    \item[(iii)] \textit{Source dipole:} The point-force and force-dipole solutions are obtained by solving the Stokes equation together with the incompressibility condition $\nabla\cdot\vec{u}=0$. In the presence of sources of solvent molecules, the Stokes equation is unchanged, but the incompressibility condition changes to $\nabla\cdot\vec{u}(\vec{r})=s(\vec{r})$, where $s(\vec{r})$ denotes the source density \cite{Graham_Book_Microhydrodynamics_2018}. While a point source is of limited relevance (it would lead to a net flow of solvent molecules entering or leaving the domain), the source dipole is an important singular solution to the Stokes equation. Its source density consists of two point sources $s_+(\vec{r})=Q\delta(\vec{r}-\vec{r}_0-(l/2)\hat{e})$ (source of solvent molecules) and $s_-(\vec{r})=-Q\delta(\vec{r}-\vec{r}_0+(l/2)\hat{e})$ (sink of solvent molecules) that are separated by a distance $l$, where $Q>0$ denotes the magnitude of the source densities. The corresponding solution to the Stokes equation in the limit $l\rightarrow 0$ reads \cite{Graham_Book_Microhydrodynamics_2018}
    \begin{equation}
        \vec{u}_{\text{SD}}(\vec{r})=\frac{Ql}{4\pi r^3}\left[3\left(\hat{e}\cdot\hat{r}\right)\hat{r}-\hat{e}\right]
        \label{eq:SourceDipole}
    \end{equation}
    and its velocity field is demonstrated in Fig.\ \ref{fig:StokesSolutionDipoles} (c).
\end{itemize}

Since self-propelled particles are force free, the simplest representation of active particles by singularity solutions of the Stokes equation is given by source and force dipoles. Examples of simulations of microswimmer models based on these singularity solutions comprise, e.g., studies of motile suspensions of active rod-like particles \cite{Saintillan_JRSocInterface_2012}, of the dynamics of a single molecule composed of microswimmers \cite{Babel_EPL_2016}, of RTPs with hydrodynamic interactions \cite{Nash_PhysRevLett_2010}, or of microswimmers near boundaries \cite{Spagnolie_JFluidMech_2012}.

To simulate microswimmers based on singularity solutions of the Stokes equation, one often models the external fluid velocity field $\vec{u}$ as a sum of all microswimmer singularity solutions and applies certain boundary conditions on the surface of each microswimmer. The velocity $\vec{v}$ of each microswimmer is then calculated using the force-free and torque-free conditions based on the stress tensor, as previously discussed for a single microswimmer, via numerical integration. To obtain self propulsion, one shifts the force or source dipole away from the center of the particles \cite{Saintillan_JRSocInterface_2012,Babel_EPL_2016}. Moreover, one can also combine the singularity solutions with numerical solvers such as the Lattice-Boltzmann method \cite{Nash_PhysRevE_2008} discussed below. Beside simulations, the force and source dipole models are used to develop continuum theories for active matter with hydrodynamic interactions, which we will discuss in the last section of this article.

\vspace{0.2cm}
\textbf{Squirmer models:} 
\\An alternative (not necessarily unrelated) approach to formulate hydrodynamically consistent models of microswimmers is to consider squirmers, i.e., spherical particles with a prescribed solvent flow along the surface (without explicitly modeling the origin of the latter) \cite{Downton_JPhysCondensMatter_2009,Goetze_PhysRevE_2010,Ishimoto_PhysRevE_2013,Zoettl_PhysRevLett_2014,Blaschke_SoftMatter_2016,Yeomans_RivNuovoCim_2017,Kuhr_SoftMatter_2017,Zoettl_EPJE_2018,Kuron_JCP_2019,Qi_PhysRveLett_2020,Zantop_SoftMatter_2020}. On the surface of the squirmer particle, the vertical fluid velocity is set to zero and the tangential surface velocity is prescribed by a series of first derivatives of Legendre polynomials, which can be used, e.g., to model the net effect of autophoresis, which leads to a slip velocity across the surface of Janus particles \cite{Popescu_EurPhysJE_2018}. The squirmer model has been used in several works, e.g., in combination with the lattice-Boltzmann method \cite{Kuron_JCP_2019,Kuron_SoftMatter_2019,Llopis_JNonNewtoniaFluidMech_2010} or multi-particle collision dynamics simulations \cite{Downton_JPhysCondensMatter_2009,Goetze_PhysRevE_2010,Zoettl_PhysRevLett_2014,Blaschke_SoftMatter_2016,Kuhr_SoftMatter_2017,Zoettl_EPJE_2018,Zantop_SoftMatter_2020,Qi_PhysRveLett_2020}.

In contrast to the ABP model and its alternatives, microswimmer models based on combinations of singularity solutions of the Stokes equation or on squirmers are momentum conserving and can correctly describe hydrodynamic interactions at large inter-particle distances for a given active system. However, they do not necessarily account for the correct hydrodynamic near-field interactions and are therefore mainly useful to model active systems at low density (squirmer models, when used to represent Janus particles, may serve as an exception, which is expected to correctly describe hydrodynamic interactions down to distances on the order of the slip length \cite{Popescu_EurPhysJE_2018}). These effective models are often used also as a starting point for continuum theories as briefly discussed further below.

\vspace{0.2cm}
\textbf{Explicit simulations of the solvent:}
\\In the following, we briefly introduce several numerical methods which are frequently used in active matter physics to explicitly determine the flow field and to simulate hydrodynamic interactions, often beyond the far-field approximation. 

\vspace{0.2cm}
\emph{Lattice-Boltzmann method:} One popular method to solve fluid dynamics problems is the lattice-Boltzmann method (LBM), where one solves the Boltzmann equation instead of the (Navier-)Stokes equation and exploits the fact that the latter equation can be derived from the former \cite{Pagonabarraga_InBook_LatticeBoltzmannModeling_2004,Krueger_Book_TheLatticeBoltzmannMethod_2017,Carenza_EPJE_2019,Desplat_CompPhysComm_2001,Ramachandran_EPJE_2006,Cates_JPhysCondensMatter_2004,Kuron_JCP_2019,Graaf_JCP_2016}. Interestingly, the Boltzmann equation is numerically often more convenient when combined with suitable approximations. It describes the particle distribution function $f(\vec{r},\vec{v},t)$, which is the density of particles with velocity $\vec{v}$ at position $\vec{r}$ and time $t$. With the so-called collision operator $\Omega(f)$, the Boltzmann equation reads \cite{Kremer_Book_IntroductionToTheBoltzmannEquation_2010,Krueger_Book_TheLatticeBoltzmannMethod_2017}
\begin{equation}
    \frac{\partial f}{\partial t} + \vec{v}\cdot\nabla_{\vec{r}}f + \frac{\vec{F}}{m}\cdot\nabla_{\vec{v}}f = \Omega(f),
    \label{eq:BoltzmannEquation}
\end{equation}
where $m$ denotes the mass of the particles and $\vec{F}$ is the external force field acting on them. The second term on the left-hand side describes advection of the particles with velocity $\vec{v}$, whereas the third term describes external forces acting on the solvent particles and affecting their velocity. The source term on the right-hand side of Eq.\ (\ref{eq:BoltzmannEquation}) describes the local redistribution of the solvent particles due to collisions. This collision operator is often approximated by $\Omega(f)=-(f-f_{\text{eq}})/\tau$, which describes the relaxation of the distribution $f$ towards the equilibrium distribution $f_{\text{eq}}$ on the time scale $\tau$ and is known as the Bhatnagar–Gross–Krook (BGK) collision operator \cite{Bhatnagar_PhysRev_1954}. In the LBM, the continuous Boltzmann equation [i.e., Eq.\ (\ref{eq:BoltzmannEquation})] is discretized in position, velocity, and time and numerically solved on a lattice with spacing $\Delta x$ at discrete times with time step $\Delta t$. The velocity $\vec{v}$ can only take discrete values $\vec{c}_i$, which are given by a discrete set $\lbrace \vec{c}_i, w_i\rbrace$ with weights $w_i$. The discretized Boltzmann equation is then solved numerically as discussed, e.g., in Ref.\ \cite{Krueger_Book_TheLatticeBoltzmannMethod_2017}. To simulate microswimmers that, e.g., create directed motion through body-shape deformations, one often describes the microswimmer surface as a set of boundary links that define a closed surface and solves the discretized Boltzmann equation together with suitable boundary conditions \cite{Krueger_Book_TheLatticeBoltzmannMethod_2017}.

\vspace{0.2cm}
\emph{Multi-particle collision dynamics:} 
\\Another popular approach to simulate the dynamics of microswimmers is based on multi-particle collision dynamics (MPCD), where, in contrast to the LBM, the solvent is represented by point-like particles which have continuous positions and velocities \cite{Gompper_InBook_MultiparticleCollisionDynamics_2009,Malevanets_InBook_MesoscopicMultiparticleCollisionModel_2004,Ruiz-Franco_JCP_2019,Winkler_CompPhysComm_2005,Winkler_JPhysCondensMatter_2004,Ripoll_EPL_2004,Lamura_EPL_2001}. To model active particles, one usually combines the MPCD method for the solvent molecules with molecular dynamics (MD) simulations of the active particles, which are coupled to the solvent and are represented either as a single particle or by a quasi-continuous distribution of particles which are connected with (time-dependent) springs and represent the surface of a (deformable) microswimmer \cite{Zoettl_ChinPhysB_2020}. The MPCD method has been used in several works to investigate, e.g., chemotactic Janus colloids \cite{Huang_NewJPhys_2017}, active particles with phoretic interactions \cite{Buyl_Nanoscale_2013}, dynamics of active particles in chemically active media \cite{Thakur_JCP_2011}, the motion of squirmers \cite{Zoettl_EPJE_2018,Qi_PhysRveLett_2020,Goetze_PhysRevE_2010}, the influence of hydrodynamic interactions on phase separation in systems of microswimmers \cite{Blaschke_SoftMatter_2016}, collective behavior of sperm cells \cite{Yang_PhysRevE_2008}, and active particles in filament networks \cite{Qiao_PhysRevRes_2020}.


\vspace{0.2cm}
\emph{Dissipative-particle dynamics:} 
\\Another coarse-grained approach to modeling the solvent is given by dissipative-particle dynamics (DPD) simulations. Here, each DPD particle represents a small solvent region and, similar to the MPCD simulations, the positions and velocities of the DPD particles take continuous values. The DPD particles interact via three types of effective forces: A weak conservative force models the soft repulsion of the solvent molecules, a dissipative force models the friction, and a random force accounts for thermal fluctuations. Knowing these forces, Newton's equation of motion is solved for the DPD particles to obtain the hydrodynamics of the solvent \cite{Fedosov_SoftMatter_2015,Hoogerbrugge_EPL_1992,Groot_JCP_1997}. This model has been adapted, e.g., to active suspensions \cite{Panchenko_PhysFluids_2018} and to model the self-propulsion of Janus colloids \cite{Eloul_PhysRevLett_2020}.

\vspace{0.2cm}
\emph{Microscopic solvent simulations:} 
\\Finally, beside the previously discussed mesoscale-simulation methods, particle-based simulations of the solvent molecules based on direct MD simulations, which allow one to resolve very small spatial and temporal scales, are possible. Nevertheless, these simulations are computationally very intense, which makes it impossible to study systems of the microscale over time scales of seconds, which are relevant to most active matter systems. Still, this explicit modeling of the solvent has been successfully used to model a self-propelled particle in a Lennard-Jones solvent \cite{Tokunaga_PhysRevE_2019}.

Overall, the LBM, MPCD, and the DPD methods are mesoscale simulation methods, which can be applied to many hydrodynamic problems in soft and active matter physics and beyond. Since the DPD method is based on particles moving in continous space, it avoids lattice artifacts and allows simulations capturing much larger length and time scales than typically possible in MD simulations. However, DPD simulations include a large number of parameters (in order to model the different forces), which have to be chosen carefully. The MPCD method, on its part, which models the net effect of individual collisions rather than accounting for every collision event, is computationally very efficient and can be efficiently parallelized. This applies also to the LBM, which numerically solves the Boltzmann equation and is well suited, e.g., for implementing complex (moving) boundaries \cite{Krueger_Book_TheLatticeBoltzmannMethod_2017}.

\section{Continuum theories of microswimmers with hydrodynamic interactions}

Based on the previously discussed effective microswimmer models, continuum theories for large ensembles of particles can be formulated which explicitly account for hydrodynamic interactions at least at low density. These theories describe wet active matter and can be formulated, e.g., based on the puller and pusher solutions of the Stokes equation. One popular approach to account for hydrodynamic far-field interactions is to write down the (overdamped) equations of motion for the position and orientation of each microswimmer, which couple with the overall fluid velocity field. The contribution of each microswimmer to the overall velocity field is modeled by singularity solutions of the Stokes equation such as force or source dipoles (which can be superimposed due to the linearity of the Stokes equation). One then derives a continuity equation for the $N$-particle probability density, which typically takes the form of a Fokker-Planck equation \cite{Heidenreich_PhysRevE_2016,Saintillan_PhysRevLett_2008,Saintillan_CRPhysique_2013,Stenhammar_PhysRevLett_2017}. From here, one can proceed in a similar way to that of microscopic theories for dry active matter in order to derive an equation of motion for the one-particle density. Since the described approach to formulate continuum theories for wet active matter is based on the singularity solutions of the Stokes equation, which only describe the far-field flow pattern of active particles, near-field hydrodynamic effects are not included in this approach. However, although complicated in practice, one can go beyond the far-field regime in principle, e.g., by using superimposed singularity solutions to represent the flow field contribution due to each swimmer or by starting with squirmer models. 

Let us finally mention that one can alternatively formulate phenomenological minimal models of wet active matter. Following a similar spirit to the case of dry active matter, these models are generic in the sense that they are largely based on considerations of symmetry, conservation laws, and dimensionality and do not refer to details such as the specific self-propulsion mechanism, which is employed by the microswimmers. One example of such a minimal model for wet active matter is given by the phenomenological active model H \cite{Tiribocchi_PhysRevLett_2015}, which accounts for momentum conservation. It is based on the active model B \cite{Wittkowski_NatComm_2014} and is closely related to the model H for equilibrium systems \cite{Hohenberg_RevModPhys_1977}. The active model H addresses the phase separation behavior of wet active matter and couples the generalized density field $\phi(\vec{r},t)$ to the velocity field $\vec{v}(\vec{r},t)$ of the solvent. The general idea is that diffusive dynamics of the active particles take place in the moving frame of the solvent and the velocity field of the solvent is given by the corresponding Navier-Stokes equation. There are also phenomenological models for specific phenomena such as bacterial turbulence, which are based on phenomenological equations to describe the fluid velocity field \cite{Wensink_PNAS_2012}.

More generally, there is a large number of alternative approaches to formulating
continuum theories for microswimmers. Readers interested in further details are referred to Refs.\ \cite{Marchetti_RevModPhys_2013,Yeomans_RivNuovoCim_2017,Lauga_RepProgPhys_2009}.

\bibliography{library_mod}

\end{document}